\documentclass[smallextended]{svjour3}       
\usepackage{graphicx}
\usepackage{url}
\usepackage[pdfpagelabels,hypertexnames=false,breaklinks=true,bookmarksopen=true,bookmarksopenlevel=2]{hyperref}
\usepackage{B2}

\usepackage{mdframed}
\DeclareFontFamily{OT1}{cmtt}{\hyphenchar \font=-1}
\DeclareFontFamily{\encodingdefault}{\ttdefault}{\hyphenchar\font=`\-}

\begin{document}

\title{Model-Driven Development of High-Assurance Active Medical Devices \thanks{The research reported in this article has been supported by the Austrian Ministry for Transport, Innovation and Technology, the Federal Ministry of Science, Research and Economy, and the Province of Upper Austria in the frame of the COMET center SCCH.} \thanks{The final publication is available at Springer via \url{https://doi.org/10.1007/s11219-015-9288-0}.
}}


\author{Atif Mashkoor 
}

%

\institute{Atif Mashkoor \at
              Software Competence Center Hagenberg GmbH  \\
              Hagenberg, Austira\\
              \email{firstname.lastname@scch.at}    
}

\date{Received: date / Accepted: date}

\maketitle

\begin{abstract}
Advanced medical devices exploit the advantages of embedded software whose development is subject to compliance with stringent requirements of standardization and certification regimes due to the critical nature of such systems. This paper presents initial results and lessons learned from an ongoing project focusing on the development of a formal model of a subsystem of a software-controlled safety-critical Active Medical Device (AMD) responsible for renal replacement therapy. The use of formal approaches for the development of AMDs is highly recommended by standards and regulations, and motivates the recent advancement of the state of the art of related methods and tools including Event-B and Rodin applied in this paper. It is expected that the presented model development approach and the specification of a high-confidence medical system will contribute to the still sparse experience base available at the disposal of the scientific and practitioner community of formal methods and software engineering. 

\end{abstract}

\section{Introduction}
\label{sec:intro}
With aging and the prevalence of noncommunicable diseases such as diabetes and hypertension, the risk of chronic kidney failure is also increasing. At the terminal phase of chronic kidney failure, renal replacement therapy is required for treatment. One of the possible forms of this therapy is hemodialysis, also known as ``artificial kidney'' treatment. It is a process in which using a pump system, the patient's blood is flowed through a special filter (dialyser) which filters out the accumulated waste to be removed together with the washer fluid at the other side of the dialyser. The machine responsible for this therapy is a classical example of an Active Medical Device (AMD).

The council directive 93/42/EEC of the European Union (EU) concerning medical devices \cite{ec93a} classifies any medical device as an AMD whose operation depends on a source of electrical energy or any source of power other than that directly generated by the human body or gravity and which acts by converting this energy. Earlier AMDs were mostly based on hardware solutions. However, lately embedded software has shown to have a determining impact on the consumer value of AMDs and their competitive differentiation. Consequently, according to the latest directive 2007/47/EC of the EU concerning medical devices \cite{ec07a}, a stand-alone software can also be considered as an AMD. The main reason of this change is that software lends itself to adaptation to individual requirements and requirements change clearly much faster than hardware.

As AMDs become more and more software-dependent, due to the immaterial nature of software, their certification becomes a crucial issue. Certification regimes have responded to this issue by proposing various related international standards such as FDA QSR, ISO 13485 \cite{iso13485}, IEC 60601-1 \cite{iec05a} and IEC 62304 \cite{iec06a}. However, instead of containing actual recommendations for techniques, tools and methods for medical software development, these standards often encourage the use of more general standards and guidelines such as IEC 61508-3 \cite{iec10a} and FDA General Principles of Software Validation \cite{fda02a} as a source for the selection of the appropriate software methods, techniques and tools. 

One of the key recommendations of almost all of the standards is to adopt formal methods for the development of software-intensive critical systems. Their use is, in fact, ``highly recommended'' at higher Safety Integrity Levels (SILs). The safety integrity of a system can be defined as the probability of a safety-related system performing the required safety function under all of the stated conditions within a stated period of time. Highly recommended means that if the mentioned technique or measure is not used, then the rationale behind this choice has to be justified during safety planning and assessment. IEC 61508-3 further states that the confidence that can be placed in the software safety requirements specification, as a basis for safe software, depends on the rigor of the techniques by which the desirable properties of the specification have been achieved. 

The overall aim of this article is to evaluate the possibility of application of a refinement-based formal approach for the improved industrial development of software-controlled safety-critical AMDs. In this case, improvement refers to increased reliability of medical systems. Specifically, we try to answer the following two research questions: 1) is the refinement-based formal approach suitable for modeling and analyzing all important elements of a complex AMD? and 2) what are advantages and challenges associated with such an approach? 

A formal model-driven development approach lets users build systems and software that are correct by construction. The combined approach of requirements modeling and analysis based on techniques such as refinement, verification and validation, and tools such as proof checkers, model checkers and animation engines results in obtaining high-assurance and trustworthy AMDs. The employed notions of formal verification and model validation are in full accordance with the related standards such as IEC 61508-3 and IEC 62304.  

To evaluate the applicability of the formal refinement-based approach, we apply it to a hemodialysis machine \cite{mashkoor15a}, that is an AMD responsible for renal replacement therapy, using the state-based formal method Event-B \cite{abrial10a} and its support platform Rodin \cite{abrial10b}.  The application is presented in this paper following the approach of conducting case study research in software engineering advocated by \cite{runeson09a}. 

The main contribution of this work is the obtained formal model that demonstrates an example of the way in which the requirements of software of modern AMDs can be rigorously specified through a chain of refinements to represent the requirements at different abstraction levels. Additionally, the paper leads to a software safety requirements specification that guarantees correctness of the addressed aspects of behavior, supports verification of the specification based on systematic analysis, avoids intrinsic specification faults, and reduces ambiguities in the specification writing process by involving customers in earlier stages of the development. It is a well-documented fact \cite{boehm88a} that the sooner an error is discovered, the lesser it costs.

The fact that the emerging role of embedded software increases the quality/cost ratio of AMDs is obviously advantageous for all the producers of such equipment. It is, however, of especially high importance for Small and Medium Enterprises (SMEs), which traditionally play a significant role in the medical device industry and which are bound to be in the forefront of innovation with their own products, as well as suppliers of larger companies. Innovation always involves higher uncertainties. However, the responsibility of SMEs in reducing the risk of the products to cause any harm is as high as that of any other company. The rigorous approach described in this paper is consequently of essential importance for SMEs in particular.

The ENISA (European Network and Information Security Agency) ad hoc working group on risk assessment and risk management report  \cite{enisa06a} already encourages SMEs to use formal paradigms for systems with medium or high criticality for businesses. The survey performed by Woodcock et al. \cite{woodcock09a} that examined industrial application of formal methods in 62 projects in the span of 25 years also yielded several interesting results, particularly for SMEs. For example, three times as many participants reported a reduction in development time as reported an increase, five times as many projects reported reductions in costs as reported an increase, and 92\% of projects reported enhanced quality compared to other techniques. The improvement was attributed to better fault detection (36\%), improved design (12\%), confidence in correctness (10\%) and better understanding (10\%). 

The paper is organized as follows. Section \ref{sec:model-development} presents the rigorous approach for modeling and analysis of high-confidence medical systems. Section \ref{sec:description} gives an overview of the selected case study. Section \ref{sec:design} discusses the case study design including objectives, the research instrument and the model development strategy. Section \ref{sec:report} first presents how the refinement-based approach has been used to model various components of a hemodialysis machine and then provides a brief account of the analysis of the presented formal model. Section \ref{sec:lessons} presents the lessons learned during the model development activity. Section \ref{sec:related} presents some related work. The paper is concluded in Section \ref{sec:conclusion}.

\section{Refinement-based model-driven development approach}
\label{sec:model-development}

The development of embedded software for AMDs is a complex process. The degree of complexity often leads to an artifact that requires a great amount of time, resources and attention to develop. However, proving its safe operation is a challenging task. While guaranteeing the absence of mistakes in a piece of software is not always possible \cite{dijkstra72a}, even the identification of their presence is not an easy task. Traditional quality assurance techniques like code reviews or test case generation are also not helpful in this case due to the critical nature of the medical domain. Additionally, the lack of domain knowledge of software engineers makes the matter worse \cite{bjorner10b}. 

We have proposed an approach where a system is synthesized using an incremental refinement process synchronizing and integrating different views and abstraction levels of the system. The process of quality assurance is embedded in the model development. Every time a requirement is specified, it goes to an internal consistency check. Once it is ensured that the requirement is specified in the right way, it is also confirmed with the stakeholders whether it indeed captures the desired behavior. The stakeholders, in this way, become part of the development process right from the start and also the chance of an error to trickle down in the later stages of the development is minimized. 

As shown by Figure \ref{fig:approach}, our approach for the development of high-assurance AMDs consists of three majors steps:

\begin{figure}
\centering
\includegraphics[width=0.75\linewidth]{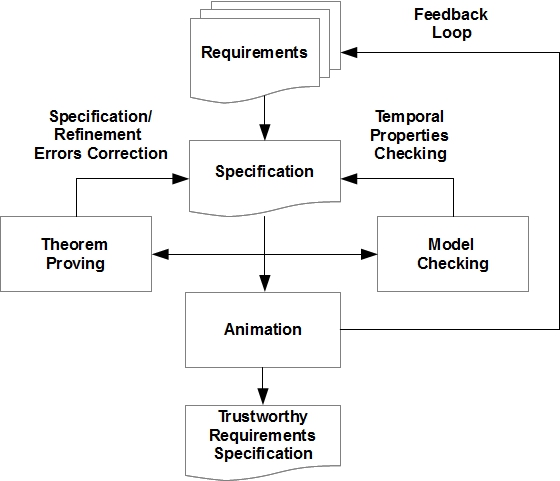}
\caption{The triptych approach of formal development of AMDs}
\label{fig:approach}
\end{figure}

\begin{enumerate}
\item formal requirements specification,
\item their verification, and
\item their validation.
\end{enumerate}

In the requirements specification step, informal user and system requirements are translated into a formal specification using a rigorous method. During this process, requirements are precisely written using mathematical and logical structures which are amenable to formal analysis to determine their correctness. 

One of the important cornerstones of the specification process is the representation of requirements at various abstraction levels using the notion of refinement. Using this technique requirements are easy to specify, analyze and implement. In this style of specification writing, requirements are added to the model in a gradual manner. Ultimately, we have a requirements model that is detailed enough to be effectively implemented.

Once the informal requirements have been translated into a formal specification, the next step is to make sure that they conform to the verification standards, i.e., requirements are consistent and verifiable. During this process it is verified that a specification conforms to some precisely expressed properties that the model is intended to fulfill such as well-definedness, invariant preservation and guard strengthening in a refinement using standard verification techniques. 

According to \cite{clarke96a}, two well established formal verification approaches are theorem proving and model checking. While the former refers to the reasoning of defined properties using a rigorous mathematical approach, the latter is the process of exploration of the whole state space of a model to verify dynamic properties. 

Both deductive theorem proving and model checking are important for proving the consistency of an AMD. While theorem proving is helpful in ensuring safety constraints of the system, model checking is effective in verifying temporal constraints of the system such as liveness and fairness properties.

Once a requirement is specified and verified, the next step to consider is its validation. It is a process where it is established by examination and provision of an objective evidence that the stakeholders' requirements have been captured correctly and completely in the requirements specification document. Verification alone is not sufficient to guarantee the correctness of the model because it does not check whether the specification documents the requirements useful for stakeholders. 

In order to make stakeholders understand the formal specification, we animate it. Animation is a process to demonstrate the fundamental operations of a specification, using a dynamic and interactive graphical display. This technique is very well-suited for making a quick mental image of the model even for non-technical domain experts. It is similar to rapid prototyping, however during animation, a specification is executed without being translated into code.

\section{Case description}
\label{sec:description}

A hemodialysis machine is used when kidneys do not perform their functions properly. i.e., removal of waste products from blood. It pumps blood from the patient's body through the arteries to the dialyser that functions as an artificial kidney or a filter. Inside the dialyser, metabolic waste products are separated from the blood. The dialyser operates as a filter that is divided into two parts by a semipermeable membrane. On one side, the patient's blood is flowing and on the other side, the dialysate. 

\begin{figure}[htpb]
\centering
\includegraphics[width=0.9\linewidth]{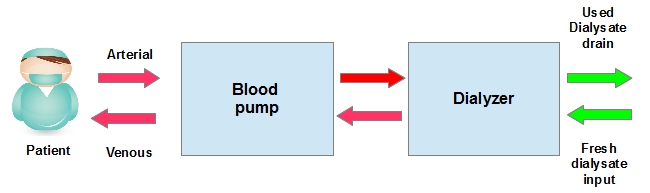}
\caption{Working principle of hemodialysis machines}
\label{fig:architecture}
\end{figure}

The dialysate, a chemical substance that is used in hemodialysis to draw fluids and toxins out of the bloodstream and to supply electrolytes and other chemicals to the bloodstream, is prepared by the hemodialysis machine for the therapy. It consists of prepared water that contains certain quantities of electrolyte and bicarbonate, depending on the individual patient's requirements. The concentrations of electrolyte and bicarbonate in the dialysate are adjusted in such a way that certain substances can be removed from the blood through convection, diffusion and osmosis, while other substances are added at the same time. This is achieved mainly by diffusive clearance through the semipermeable membrane of the dialyser. The dialysate transports the metabolic waste products from the dialyser into the discharge line. The cleaned blood is then recycled back to the patient through the venous access. The working principle of the hemodialysis machine is depicted by Figure \ref{fig:architecture}.

The detailed description of the case study is available in \cite{mashkoor15c}.

\section{Case study design}
\label{sec:design}

\subsection{Objectives}
\label{subsec:objectives}

The objective of this study is to evaluate the applicability of  refinement-based processes for improved embedded software development of active medical devices. We want to determine whether this approach results in an increased reliability of a medical device (demonstrable by proof) while having an effective grasp on the notion of verification and validation throughout its development life cycle. With the employed rigorous model-based approach, the goal is expected to be achieved. Thus, we wanted to investigate advantages and challenges of the proposed approach, and evaluate the effects of implementing it.

The concrete research questions for this study are as follows:
\begin{enumerate}
\item Is the formal model-based approach sufficient for modeling all elements of a complex active medical device?
\item What are advantages and challenges associated with this approach? 
\end{enumerate}

\subsection{Research instrument}
\label{subsec:instrument}

We use the formal method Event-B \cite{abrial10a} for model development. It is based on Zermelo-Fraenkel set theory with the axiom of choice. It is the successor of the B method \cite{abrial96a} for the development of complex reactive systems. We have chosen this method for the formal development of AMDs because of its ability to represent systems at various abstraction levels using its refinement mechanism, easy to use modeling notation and the extensive tool support.

\subsubsection*{The modeling language}

A typical Event-B model is composed of two constructs: machines and contexts. Machines define the dynamic behavior of the model. A typical Event-B machine includes: 
\begin{itemize}
\item variables, which define the state space of the machine and can be expressed using natural numbers, integers, real numbers, boolean, sets, relations, functions or any other set-theoretical construct,
\item invariants, which are used either to type variables or to constrain the state space of the machine,
\item variants, which are used to define the convergence property of events, i.e., they can be triggered only for a finite number of times. They are defined using either a natural number or a finite set, 
\item and events, which describe state transitions. An event is defined as a binary relation composed of guards and actions. A guard is a predicate and all the guards together construct the domain of the corresponding relation. An action is an assignment statement to a state variable and is achieved by a generalized substitution. Combined together, all actions form the range of the corresponding relation. The actions of a particular event are executed simultaneously and non-deterministically.    
\end{itemize}

Contexts define the static elements of a model. They contain carrier sets, constants, axioms and theorems. Carrier sets are used to define types. Axioms are used to constrain carrier sets and constants. Theorems define properties that are derived from axioms.  

\subsubsection*{The refinement process}

The Event-B method uses the refinement process to transform an abstract specification into a concrete one. An Event-B model can be refined in several ways: 
\begin{itemize}
\item new variables and invariants can be introduced and existing invariants can be strengthened,
\item existing events can be refined to include and preserve new and existing variables and invariants respectively,
\item existing events can be split into several new events, and,
\item completely new events can be introduced to the model.
\end{itemize}

A machine can be refined into another machine which then contains a more detailed description of the model. A machine can see several contexts, i.e., use the constants and axioms they contain. A context can also be further refined into one or more contexts and can be seen by several machines.

\subsubsection*{The Rodin toolset}
Rodin \cite{abrial10b} is the tool that supports modeling and analysis in the Event-B method. Rodin is built upon the Eclipse platform and is extensible by plug-ins such as the model checking and animation plug-in ProB \cite{leuschel03a}. The main tasks that are supported by Rodin are: 
\begin{itemize}
\item specification of machines and contexts,
\item their refinement, and
\item their consistency checking by automatically generating Proof Obligations (POs).
\end{itemize}

Proofs can be discharged either automatically, with the help of third-party theorem provers, or interactively. Animation can be achieved by making scenarios and then making sure that events are being fired in the desired order.

\subsection{Model development and refinement strategy}
\label{subsec:strategy}

During the requirements modeling process, following the advice of \cite{mashkoor11c} to take small refinement steps, we chose to introduce one requirement per refinement level. Every refinement step introduces a new monolithic requirement of the corresponding component into the model. The static data related to requirements is modeled in contexts and the general behavior of the system is presented in machines using events. The safety requirements are specified as machine invariants. 

In order to improve the legibility of our requirements specification, as originally proposed by \cite{mashkoor10b}, we have classified our axioms into three groups: technical axioms, typing axioms and property axioms. This practice helps in distilling the actual software requirements from technical expressions.

As the Event-B method lacks the explicit notion of time, we have used the timing pattern for Event-B proposed by \cite{cansell07a}. In this technique, $\mathbb{N}$ is used to model the notion of time. 

In order to conserve space, the shown refinements contain only the newly introduced information instead of the complete model.

\section{Case study report}
\label{sec:report}

\subsection{Model development}

The components of the hemodialysis machine we have chosen to demonstrate as the case study in this paper belong to three different categories: 
\begin{enumerate}
\item The first one is responsible for connecting the patient to the machine.
\item The second is responsible for monitoring the blood flow at set rates from a patient's arterial access.
\item The last one is responsible for maintaining the temperature of the dialysate.
\end{enumerate}

\subsubsection{The patient connection component}

The {\it connect patient} component is responsible for establishing a connection between a patient and a machine.

\subsubsection*{Abstract model}

The abstract model is comprised of the following requirement:

\begin{mdframed}
The software shall monitor the blood flow in the Extra-corporeal Blood Circuit (EBC) and if no flow is detected, then the software shall stop the blood pump and execute an alarm signal.
\end{mdframed}

We first initiate a context ({\tt Context CCP0}), as shown in Figure \ref{fig:CCP0}, that contains the static data to specify this requirement. It has two sets: {\tt BloodPumpingValues} which models the state of the blood pumping process ({\tt BPStarted} or {\tt BPStopped}), and {\tt Alarms}, which contains different types of alarms of the system. The alarm {\tt ALM382} is related to this particular requirement. The constant {\tt Null} defines a state where no alarm has been triggered.

\begin{figure}[htbp]
\begin{center}
  \begin{minipage}{0.7\linewidth}
\begin{lstlisting}[frame=single]
CONTEXT
  CCP0
SETS
  BloodPumpingValues, Alarms
CONSTANTS
  BPStarted, BPStopped, ALM382, Null
AXIOMS
  typ1 partition(BloodPumpingValues, {BPStarted}, {BPStopped})
  typ2 partition(Alarms, {ALM382}, {Null})
END
\end{lstlisting}
  \end{minipage}
\end{center}
\caption{Context CCP0}
\label{fig:CCP0}
\end{figure}

The corresponding machine {\tt MCP0} specifies the aforementioned requirement as shown by {\tt inv4} in Figure \ref{fig:MCP0}, i.e., if no flow is detected, then the software shall stop the blood pump and execute the related alarm. The first three invariants of the machine specify the typing of the variables. The last invariant states that the blood pumping process also implicates the blood flowing process.

In order to capture the behavior of the system, the following events have been introduced to the machine. 

\begin{itemize}
\item The event {\tt INITIALISATION} is the default event to initialize the values of newly introduced variables. 
\item The event {\tt startBloodPumping} is trivial as it is used to start the blood pumping process.
\item The event {\tt stopBloodPumping} is also trivial as it just checks if the blood pumping process is already started. If so, it stops this process and sets the blood flow state as false.
\item The event {\tt bloodFlowMonitoring} actually specifies the monitoring process of blood flow. If no flow is detected, then the action part of the event stops the blood flow pumping process and triggers the related alarm.
\end{itemize}

\begin{figure}[htb]
\begin{center}
  \begin{minipage}{0.6\linewidth}
\begin{lstlisting}[frame=single]
MACHINE
  MCP0
SEES
  CCP0 
VARIABLES 
  bloodFlow, alarm, bloodPumping
INVARIANTS
  inv1 bloodFlow : BOOL 
  inv2 alarm : Alarms 
  inv3 bloodPumping : bloodPumpingValues 
  inv4 bloodPumping = BPStarted & bloodFlow = FALSE =>
       bloodPumping = BPStopped & alarm = ALM382  
  inv5 bloodPumping = BPStarted => bloodFlow = TRUE 
EVENTS
  Event INITIALISATION 
    Then
      act1 bloodFlow := FALSE
      act2 alarm := Null
      act3 bloodPumping := BPStopped
  End
  Event startBloodPumping 
    Where
      grd1 bloodPumping = BPStopped 
    Then
      act1 bloodPumping := BPStarted
      act2 bloodFlow := TRUE 
  End
  Event stopBloodPumping 
    Where
      grd1 bloodPumping = BPStarted 
    Then
      act1 bloodFlow := FALSE 
      act2 bloodPumping := BPStopped 
  End
  Event bloodFlowMonitoring 
    Where
      grd1 bloodFlow = FALSE
      grd2 bloodPumping = BPStarted 
    Then
      act1 bloodPumping := BPStopped
      act2 alarm := ALM382
  End
END
\end{lstlisting}
  \end{minipage}
\end{center}
\caption{Machine MCP0}
\label{fig:MCP0}
\end{figure}

\subsubsection*{First refinement}

The first refinement includes the following requirement into the model:

\begin{mdframed}
The software shall monitor the filling blood volume of the EBC and if the blood volume exceeds 400 ml then the software shall stop the blood pump and execute an alarm signal.
\end{mdframed}

To model this requirement, the context {\tt CCP0} is extended to {\tt CCP1} which simply introduces the new alarm type, i.e., {\tt ALM344}, related to this requirement.

The corresponding machine {\tt MCP1} at this level specifies the requirement using the following invariant:

\[\begin{array}{c}
bloodPumping = BPStarted \wedge bloodFlow = TRUE \wedge \\ fillingBloodVolume > 400 \Rightarrow \\
bloodPumping = BPStopped \wedge alarm = ALM344
\end{array} \]

A new variable {\tt fillingBloodVolume} is introduced to keep track of the filling volume along with a new monitoring event {\tt fillingBloodVolumeMonitoring} that is shown in Figure \ref{fig:fillingBloodVolumeMonitoring}.

\begin{figure}[htbp]
\begin{center}
  \begin{minipage}{0.4\linewidth}
\begin{lstlisting}[frame=single]
Event fillingBloodVolumeMonitoring
 Where
  grd1 fillingBloodVolume > 400
  grd2 bloodPumping = BPStarted
  grd3 bloodFlow = TRUE
 THEN
  act1 bloodPumping := BPStopped 
  act2 alarm := ALM344  
  act3 bloodFlow := FALSE
END
\end{lstlisting}
  \end{minipage}
\end{center}
\caption{Event fillingBloodVolumeMonitoring}
\label{fig:fillingBloodVolumeMonitoring}
\end{figure}

\subsubsection*{Second refinement}

The second refinement includes the following requirement into the model:

\begin{mdframed}
The software shall use a timeout of 310 seconds after the first start of the blood pump. After this timeout the software shall change to the therapy mode.
\end{mdframed}

The context {\tt CCP2} extends the context {\tt CCP1} by adding a new set {\tt SoftwareMode} that is comprised of the following modes: {\tt Therapy, TherapySelection, Preparation, EndOfTherapy, Disinfection}.

The corresponding machine {\tt MCP2} at this level specifies the current requirement by the following invariant. 

\[\begin{array}{c}
bloodPumping = BPStarted \wedge bloodPumpingTime > 310 \wedge \\ softwareMode = Preparation \Rightarrow softwareMode = Therapy
\end{array} \]

Two new variables are added to the machine. The variable {\tt bloodPumpingTime} simulates the tick of the clock for the blood pumping process and the variable {\tt softwareMode} represents the current mode of the software.

Two new events are also introduced to model the functionality. Event {\tt tick}, shown in Figure \ref{fig:tick}, acts as a clock that ticks at regular intervals. Event {\tt changeMode}, shown in Figure \ref{fig:changeMode}, is the event that is responsible for changing the mode of the software from preparation to therapy after 310 seconds.

\begin{figure}[htbp]
{
\begin{minipage}{0.55\linewidth}
\begin{lstlisting}[frame=single]
Event tick
 Where
  grd1 bloodPumping = BPStarted
  grd2 bloodPumpingTime <= 310
  
 Then
  act1 bloodPumpingTime := bloodPumpingTime + 1
End
\end{lstlisting}
\caption{Event tick}
\label{fig:tick}
\end{minipage}
\hfill
\begin{minipage}{0.4\linewidth}
\begin{lstlisting}[frame=single]
Event changeMode
 Where
  grd1 bloodPumping = BPStarted
  grd2 bloodPumpingTime > 310
  grd3 softwareMode = Preparation
 Then
  act1 softwareMode := Therapy
End
\end{lstlisting}
\caption{Event changeMode}
\label{fig:changeMode}
\end{minipage}
\hfill}
\end{figure}

\subsubsection*{Third refinement}

The third refinement includes the following requirement into the model:

\begin{mdframed}
The software shall monitor the blood flow direction and if the reverse direction is detected, then the software shall stop the blood pump and execute an alarm signal.
\end{mdframed}

The context {\tt CCP3} extends the context {\tt CCP2} by adding a new set {\tt BloodFlowDirectionValues} that are either {\tt Forward} or {\tt Backward}. The new related alarm {\tt ALM737} is also added to the context.

The corresponding machine {\tt MCP3} introduces a new variable {\tt bloodFlowDirection} to monitor the direction of the blood flow. The requirement is then specified using the following invariant: 

\[\begin{array}{c}
bloodPumping = BPStarted \wedge bloodFlowDirection = Backward \Rightarrow \\ bloodPumping = BPStopped \wedge alarm = ALM737
\end{array} \]

The newly introduced event {\tt bloodFlowDirectionMonitoring}, shown in Figure \ref{fig:bloodFlowDirectionMonitoring}, defines the monitoring event that checks the current blood flow direction and if it is going backward, it immediately stops the blood pumping process and triggers the related alarm.

\begin{figure}[htbp]
\begin{center}
  \begin{minipage}{0.4\linewidth}
\begin{lstlisting}[frame=single]
Event bloodFlowDirectionMonitoring
 Where
  grd1 bloodFlowDirection = Backward
  grd2 bloodPumping = BPStarted
 THEN
  act1 bloodPumping := BPStopped 
  act2 alarm := ALM737  
 END
\end{lstlisting}
  \end{minipage}
\end{center}
\caption{Event bloodFlowDirectionMonitoring}
\label{fig:bloodFlowDirectionMonitoring}
\end{figure}

\subsubsection{The blood pumping component}

The {\it blood pumping} component is responsible for blood flow at set rates from the arterial access of patients through the dialyser to their venous access. 

\subsubsection*{Abstract model}

The abstract model of the {\it blood pumping} component contains the following requirement: 

\begin{mdframed}
The software shall monitor the blood flow in the EBC and if no flow is detected for more than 120 seconds, then the software shall stop the blood pump and execute an alarm signal.
\end{mdframed}

Like the previous component, we first initiated a context ({\tt Context CBP0}) that is exactly the same as the context {\tt Context CCP0} shown in Figure \ref{fig:CCP0}; the same alarm is triggered in both cases.

The corresponding machine {\tt MBP0} introduces a new variable {\tt noFlowDetectionTime} that is used to simulate the behavior of a clock related to the blood flow. The requirement is then specified by the following invariant:

\[\begin{array}{c}
bloodPumping = BPStarted \wedge noFlowDetectionTime > 120 \Rightarrow \\ bloodPumping = BPStopped \wedge alarm = ALM382
\end{array} \]

The monitoring event {\tt noFlowMonitoring}, shown in Figure \ref{fig:noFlowMonitoring}, specifies the behavior of the model in case that no blood flow is detected for more than 120 seconds. The blood pumping process is then stopped, the related alarm is triggered and the related clock is reset.

\begin{figure}[htbp]
\begin{center}
\begin{minipage}{0.37\linewidth}
\begin{lstlisting}[frame=single]
Event noFlowMonitoring
 Where
  grd1 bloodPumping = BPStarted
  grd2 noFlowDetectionTime >= 120
  grd3 bloodFlow = FALSE
 Then
  act1 bloodPumping := BPStopped
  act2 alarm := ALM382
  act3 noFlowDetectionTime := 0
End
\end{lstlisting}
\caption{Event noFlowMonitoring}
\label{fig:noFlowMonitoring}
\end{minipage}
\end{center}
\end{figure}

\subsubsection*{First refinement}

The first refinement introduces the following requirement into the model:

\begin{mdframed}
If the system is not in bypass then the software shall monitor the blood in the EBC and if the actual blood flow is less than 70\% of the set blood flow, then the software shall execute an alarm signal.
\end{mdframed}

The context {\tt CBP0} is extended into {\tt CBP1} which simply introduces the new alarm type {\tt ALM755} related to this particular requirement along with a constant {\tt SetBloodFlow} that sets the desired amount of blood flow in the system.

The corresponding machine {\tt MBP1} at this level introduces two new variables into the specification. The variable {\tt actualBloodFlow} represents the current amount of flowing blood and the variable {\tt bypass} represents the bypass state of the system, i.e., the bypass valve is closed or not. The following invariant is introduced for modeling the related requirement: 

\[\begin{array}{c}
bypass = FALSE \wedge actualBloodFlow < ((7 \ast SetBloodFlow) \div 10) \Rightarrow \\ alarm = ALM755
\end{array} \]

This refinement also introduces the monitoring event  {\tt lessBloodFlowMonitoring} into the model that is shown by Figure \ref{fig:lessBloodFlowMonitoring}. It makes sure that the actual amount of flowing blood does not fall short of its set amount. If this happens, then it triggers the related alarm.

\begin{figure}[htbp]
\begin{center}
  \begin{minipage}{0.5\linewidth}
\begin{lstlisting}[frame=single]
Event lessBloodFlowMonitoring
 Where
  grd1 bypass = FALSE
  grd2 bloodPumping = BPStarted
  grd3 actualBloodFlow < ((7*SetBloodFlow)/10)
 THEN
  alarm := ALM755
 END
\end{lstlisting}
  \end{minipage}
\end{center}
\caption{Event lessBloodFlowMonitoring}
\label{fig:lessBloodFlowMonitoring}
\end{figure}

\subsubsection*{Second refinement}

The second refinement includes the following requirement into the model:
\begin{mdframed}
The software shall monitor the rotation direction of the blood flow pump and if the software detects that the blood flow pump rotates backwards, then the software shall stop the blood flow pump and execute an alarm signal.
\end{mdframed}

This requirement is pretty much same as the requirement which we modeled in the third refinement of the {\it connect patient} component. Hence, we simply copy the same formalization into the model.

\subsubsection{The temperature monitoring component}

This component is responsible for monitoring the temperature of the dialysate delivered to the dialyser. The dialysate preparation consists in mixing the heated and degassed dialysate water with other fluid concentrates. 

\subsubsection*{Abstract model}

The abstract model of the {\it temperature monitoring} component contains the following requirement: 

\begin{mdframed}
If the system is in the preparation mode and performs priming or rinsing or if the system is in the therapy mode and if the dialysate temperature exceeds the maximum temperature of 41$^{\circ}$C, then the software shall disconnect the dialyser from the dialysate and execute an alarm signal.
\end{mdframed}

Like the previous components, we first initiate a context ({\tt Context CTM0}) that is shown in Figure \ref{fig:CTM0}. The first five axioms of the context are typing axioms. The last one is specified to initialize the starting state of the dialyser fluids, i.e., all the fluids are disconnected at the time of initialization.

\begin{figure}[htbp]
\begin{center}
  \begin{minipage}{0.99\linewidth}
\begin{lstlisting}[frame=single]
CONTEXT
 CTM0
SETS
 Operations, DialyserStates, Fluids, Alarms, SoftwareModes 
CONSTANTS
 Priming, Rinsing, DialyserConnected, DialyserDisconnected, ALM377, Default, ALM639, Dialysate, 
 StartingDialysingState, Therapy, TherapySelection, Preparation, EndOfTherapy, Disinfection, Null 
AXIOMS
 typ1 partition(Operations, {Priming}, {Rinsing}, {Default})
 typ2 partition(DialyserStates, {DialyserConnected}, {DialyserDisconnected})
 typ3 partition(Alarms, {Null}, {ALM377}, {ALM639})
 typ4 partition(Fluids, {Dialysate})
 typ5 partition(SoftwareModes, {Therapy}, {TherapySelection}, {Preparation}, {EndOfTherapy}, 
      {Disinfection})
 tec1 StartingDialysingState : Fluids --> {DialyserDisconnected}
END
\end{lstlisting}
  \end{minipage}
\end{center}
\caption{Context CTM0}
\label{fig:CTM0}
\end{figure}

The corresponding machine {\tt MTM0} is shown in Figure \ref{fig:MTM0}. It contains several variables whose typing is given by first five invariants. Invariants {\tt inv6} and {\tt inv7} specify the related requirement. The requirement is split into two invariants because each software mode executes a different alarm if the fluid temperature exceeds the limit of 41$^{\circ}$C.

Correspondingly, we specify two different events to capture the behavior of the system. If software is in the preparation mode and the temperature of the dialysate rises to more than 41$^{\circ}$C during the operation, then the event {\tt disconnectDialyserPreparation} is triggered. If the software is in the therapy mode while the same thing happens, then the event {\tt disconnectDialyserTherapy} is triggered.

\begin{figure}[htbp]
\begin{center}
  \begin{minipage}{0.99\linewidth}
\begin{lstlisting}[frame=single]
MACHINE
 MTM0
SEES
 CTM0 
VARIABLES 
 dialyserState, dialysateTemperature, operation, softwareMode, alarm
INVARIANTS
 inv1 dialyserState : Fluids --> DialyserStates
 inv2 dialysateTemperature : NAT
 inv3 operation : Operations
 inv4 softwareMode : SoftwareModes
 inv5 alarm : Alarms
 inv6 softwareMode = Preparation & (operation = Priming or operation = Rinsing) & 
      dialysateTemperature > 41 => dialyserState = {Dialysate |-> DialyserDisconnected} & 
      alarm = ALM377
 inv7 softwareMode = Therapy & dialysateTemperature > 41 =>
      dialyserState = {Dialysate |-> DialyserDisconnected} & alarm = ALM639
EVENTS
 Event Initialization 
  Then
   act1 alarm := Null
   act2 operation := Default
   act3 dialyserState := StartingDialysingState
   act4 dialysateTemperature := 0
   act5 softwareMode := Preparation
  End
 Event disconnectDialyserPreparation
  Where
   grd1 softwareMode = Preparation
   grd2 dialysateTemperature > 41
   grd3 operation = Priming or operation = Rinsing
   grd4 dialyserState = {Dialysate |-> DialyserConnected}
  Then
   act1 dialyserState := {Dialysate |-> DialyserDisconnected}
   act2 alarm := ALM377
  End
 Event disconnectDialyserTherapy
  Where
   grd1 softwareMode = Therapy
   grd2 dialysateTemperature > 41
   grd3 dialyserState = {Dialysate |-> DialyserConnected}
  Then
   act1 dialyserState := {Dialysate |-> DialyserDisconnected}
   act2 alarm := ALM639
  End
END
\end{lstlisting}
  \end{minipage}
\end{center}
\caption{Machine MTM0}
\label{fig:MTM0}
\end{figure}

\subsubsection*{First refinement}

The first refinement of the {\it temperature monitoring} component contains the following requirement:

\begin{mdframed}
If the system is in the therapy mode and if the dialysate temperature falls below the minimum temperature of 33$^{\circ}$C, then the software shall disconnect the dialyser from the dialysate and execute an alarm signal.
\end{mdframed}

The context {\tt CTM0} is extended to {\tt CTM1} which simply introduces the new alarm type {\tt ALM757} related to this particular requirement.

The corresponding machine {\tt MTM1} introduces the following invariant:

\[\begin{array}{c}
softwareMode = Therapy \wedge dialysateTemperature < 33 \Rightarrow \\ dialyserState = \{Dialysate \mapsto DialyserDisconnected\} \wedge \\ alarm = ALM757
\end{array} \]

The model is further strengthened by the introduction of the event {\tt disconnectDialyserTherapyII} as shown in Figure \ref{fig:disconnectDialyserTherapyII}. It states that if the dialysate temperature drops from 33$^{\circ}$C during the therapy mode, then the software should disconnect the dialysate from the dialyser and trigger the related alarm. 

\begin{figure}[htbp]
\begin{center}
  \begin{minipage}{0.7\linewidth}
\begin{lstlisting}[frame=single]
Event disconnectDialyserTherapyII
 Where
  grd1 softwareMode = Therapy
  grd2 dialysateTemperature < 33
  grd3 dialyserState = {Dialysate |-> DialyserConnected}
 Then
  act1 dialyserState := {Dialysate |-> DialyserDisconnected}
  act2 alarm := ALM757
End
\end{lstlisting}
  \end{minipage}
\end{center}
\caption{Event disconnectDialyserTherapyII}
\label{fig:disconnectDialyserTherapyII}
\end{figure}

\subsubsection*{Second refinement}

The second refinement of the {\it temperature monitoring} component contains the following requirement:

\begin{mdframed}
If the system is in the preparation mode and performs priming or rinsing or if the system is in the therapy mode and the HemoDiaFiltration (HDF) option is available and if the Substitution Fluid (SF) valve is opened and if the dialysate temperature exceeds 42$^{\circ}$C, then the software shall disconnect the dialyser from the dialysate and the EBC from the SF, request bypass and execute an alarm signal.
\end{mdframed}

The context {\tt CTM1} is extended to {\tt CTM2} as shown in Figure \ref{fig:CTM2}. It introduces three sets, i.e.,  {\tt HDFValues}, {\tt SFValveValues} and {\tt EBCStates}, and various constants. Axioms {\tt typ1} to {\tt typ4} assign the constants to the sets. The last axiom {\tt tec1} defines the initial state of the EBC. 

\begin{figure}[htbp]
\begin{center}
  \begin{minipage}{0.75\linewidth}
\begin{lstlisting}[frame=single]
CONTEXT
 CTM2
EXTENDS
 CTM1
SETS
 HDFValues, SFValveValues,  EBCStates 
CONSTANTS
 Online, Offline, Opened, Closed, SubstituitionFluid,  CircuitConnected, 
 CircuitDisconnected, StartingEBCState 
AXIOMS
 typ1 partition(HDFValues, {Online}, {Offline})
 typ2 partition(SFValues, {Opened}, {Closed})
 typ3 SubstituitionFluid : Fluids
 typ4 partition(EBCStates, {CircuitConnected}, {CircuitDisconnected})
 tec1 StartingEBCState : Fluids --> {CircuitDisconnected}
End
\end{lstlisting}
  \end{minipage}
\end{center}
\caption{Context CTM2}
\label{fig:CTM2}
\end{figure}

The corresponding machine defines several new variables. The variable {\tt optionHDF} is defined to check the availability of the HDF option\footnote{HemoDiaFiltration (HDF) is a process in which a high rate of ultrafiltration is used for dialysis.}. The variable {\tt SFValveValue} determines the opening and closing of the substitution fluid valve. The variable {\tt EBCState} determines which fluid is currently connected to the EBC. The variable {\tt dialysateBypass} determines whether the bypass option for the dialysate has been selected or not. The following invariants are also included in the model to state the requirement:

\[\begin{array}{c}
softwareMode = Preparation \wedge (operation = Priming \vee operation = Rinsing) \\
\wedge optionHDF = Online \wedge SFValveValue = Opened \wedge \\ dialysateTemperature > 42 \Rightarrow \\ dialyserState = \{Dialysate \mapsto DialyserDisconnected\} \wedge \\
EBCState = \{SubstituitionFluid \mapsto CircuitDisconnected\} \wedge \\
dialysateBypass = TRUE \wedge alarm = ALM377
\end{array} \]

\[\begin{array}{c}
softwareMode = Therapy \wedge optionHDF = Online \wedge SFValveValue = Opened \wedge \\ dialysateTemperature > 42 \Rightarrow \\ dialyserState = \{Dialysate \mapsto DialyserDisconnected\} \wedge \\ EBCState = \{SubstituitionFluid \mapsto CircuitDisconnected\} \wedge \\
dialysateBypass = TRUE \wedge alarm = ALM639
\end{array} \]

Two new events are also added to the model. The event {\tt disconnectDialyserTherapyIII}, shown in Figure \ref{fig:disconnectDialyserTherapyIII}, demonstrates how and when to disconnect the dialyser during the therapy mode when the temperature exceeds the specified limit. The event {\tt disconnectDialyserPreparationII}, shown in Figure \ref{fig:disconnectDialyserPreparationII}, demonstrates the same requirement for the preparation mode.

\begin{figure}[htbp]
{
\begin{minipage}{0.48\linewidth}
\begin{lstlisting}[frame=single]
Event disconnectDialyserTherapyIII
 Where
  grd1 softwareMode = Therapy
  grd2 optionHDF = Online
  grd3 SFValveValue = Opened
  grd4 dialysateTemperature > 42
  
 Then
  act1 dialyserState := 
      {Dialysate |-> DialyserDisconnected}
  act2 EBCState := 
      {SubstituitionFluid |-> CircuitDisconnected}
  act3 alarm := ALM639
  act4 dialysateBypass := TRUE
End
\end{lstlisting}
\caption{Event disconnectDialyserTherapyIII}
\label{fig:disconnectDialyserTherapyIII}
\end{minipage}
\hfill
\begin{minipage}{0.48\linewidth}
\begin{lstlisting}[frame=single]
Event disconnectDialyserPreparationII
 Where
  grd1 softwareMode = Preparation
  grd2 operation = Priming or operation = Rinsing
  grd3 optionHDF = Online
  grd4 SFValveValue = Opened
  grd5 dialysateTemperature > 42
 Then
  act1 dialyserState := 
      {Dialysate |-> DialyserDisconnected}
  act2 EBCState := 
      {SubstituitionFluid |-> CircuitDisconnected}
  act3 alarm := ALM377
  act4 dialysateBypass := TRUE
End
\end{lstlisting}
\caption{Event disconnectDialyserPreparationII}
\label{fig:disconnectDialyserPreparationII}
\end{minipage}
\hfill}
\end{figure}

\subsubsection*{Third refinement}

The third refinement of the {\it temperature monitoring} component contains the following requirement:

\begin{mdframed}
If the system is in the therapy mode and the HDF option is available and if the SF valve is opened and if the dialysate temperature falls below 33$^{\circ}$C, then the software shall disconnect the dialyser from the dialysate and the EBC from the SF, request bypass and execute an alarm signal.
\end{mdframed}

No new context is added to the model as no new static piece of information is introduced by this requirement. 

The corresponding machine introduces the following invariant:

\[\begin{array}{c}
softwareMode = Therapy \wedge optionHDF = Online \wedge SFValveValue = Opened \wedge \\ dialysateTemperature < 33 \Rightarrow \\ dialyserState = \{Dialysate \mapsto DialyserDisconnected\} \wedge \\ EBCState = \{SubstituitionFluid \mapsto CircuitDisconnected\} \wedge \\ dialysateBypass = TRUE \wedge alarm = ALM757
\end{array} \]

To capture the behavior specified by the requirement, a new event {\tt disconnectDialyserTherapyIV} as shown in Figure \ref{fig:disconnectDialyserTherapyIV} is added to the model. It specifies how the system should react in the therapy mode when the HDF option is enabled, the SF valve is open and the temperature of the dialysate drops below the minimum threshold of 33$^{\circ}$C. In this case both the dialysate and the SF are disconnected, dialysate bypass is enabled and the related alarm is triggered.   

\begin{figure}[htbp]
\begin{center}
  \begin{minipage}{0.65\linewidth}
\begin{lstlisting}[frame=single]
Event disconnectDialyserTherapyIV
 Where
  grd1 softwareMode = Therapy
  grd2 optionHDF = Online
  grd3 SFValveValue = Opened
  grd4 dialysateTemperature < 33
 Then
  act1 dialyserState := {Dialysate |->  DialyserDisconnected}
  act2 EBCState := {SubstituitionFluid |-> CircuitDisconnected}
  act3 dialysateBypass := TRUE
  act4 alarm := ALM757
End
\end{lstlisting}
  \end{minipage}
\caption{Event disconnectDialyserTherapyIV}
\label{fig:disconnectDialyserTherapyIV}
\end{center}
\end{figure}
\subsection{Formal analysis of the model}
\label{subsec:analysis}
A model is considered to be formally correct when it is both verified and validated. Verification of a model is achieved when it is proved that it is free from specification errors and inconsistencies. This is usually done either through the system of POs or through model checking. A proved specification ensures that it is consistent, well-defined and its events preserve its invariants. However, proving a refinement requires to  prove that concrete events maintain invariants of the abstract model, maintain abstraction invariants, and, when appropriate, decrease variants monotonically. Using model checking, we make sure that states of a model are reachable, its formulas are satisfiable and it does not contain deadlocks. 

For our model, Rodin generated three kinds of POs: 1) invariants preservation, 2) well-definedness of guards and invariants, and 3) equality of a preserved variable.   

Invariants preservation relates to the condition that each variable affected by the assignment statement must preserve the invariant. For example, the event {\tt stopBloodPumping} of machine {\tt MCP0}, shown in Figure \ref{fig:MCP0}, using its guard and both actions ensures that the related invariants {\tt inv2} and {\tt inv5} of the machine are preserved.

The notion of well-definedness relates to the condition which leads to safe evaluation of an expression. For example, the invariant of machine {\tt MCP1} states a condition where the variable {\tt actualBloodFlow} of type $\mathbb{N}$ is compared to an expression of type $\mathbb{R}$. However, as the value assigned to {\tt actualBloodFlow} is always of type $\mathbb{N}$, well-definedness is provable.

Equality of a preserved variable amounts to proving that if a variable is present in both the abstract as well as the concrete machine and an event of the concrete machine assigns a (new) value to this variable, then it must be proven that this value is consistent with the previous one. For example, the variable {\tt alarm} in all the machines is assigned with a new alarm, however all the alarms belong to the same type, i.e., {\tt Alarms}.

Table 1 expresses the proof statistics for our formal development using the Rodin platform. These statistics measure the total number of generated POs, automatically discharged POs by the Rodin platform, and manually discharged POs. The development of components of the dialysis machine resulted in 106 POs, out of which 105 were discharged automatically. Only one PO required manual interaction (only a few clicks). The employed approach of incremental development helped us to achieve a high degree of automatically discharged POs.

\begin{table}
\begin{center}
\begin{tabular}{|c|c|c|c|} \hline
& {\bf Total} & {\bf Automatic} & {\bf Interactive} \\ \hline
${\bf Patient}$ ${\bf connection}$ &  {\bf 35} & {\bf 35} & {\bf 0} \\ \hline 
 $MCP0$ &  8 &  8 & 0 \\ \hline
 $MCP1$ &  9 &  9 & 0 \\ \hline
 $MCP2$ &  10 &  10 & 0 \\ \hline
 $MCP3$ &  8 &  8 & 0 \\ \hline
 ${\bf Blood}$ ${\bf pumping}$ &  {\bf 25} & {\bf 24} & {\bf 1} \\ \hline
 $MBP0$ &  9 &  9 & 0 \\ \hline
 $MBP1$ &  8 &  7 & 1 \\ \hline
 $MBP2$ &  8 &  8 & 0 \\ \hline
 ${\bf Temperature}$ ${\bf monitoring}$ &  {\bf 46} & {\bf 46} & {\bf 0} \\ \hline 
 $MTM0$ &  10 & 10 & 0 \\ \hline
 $MTM1$ &  6 &  6 & 0 \\ \hline
 $MTM2$ &  19 &  19 & 0 \\ \hline
 $MTM3$ &  11 &  11 & 0 \\ \hline
 ${\bf Total}$ &  {\bf 106} & {\bf 105} & {\bf 1} \\ \hline 
\end{tabular}
\end{center}
\label{tab:stats}
 \caption{Proof statistics}
\end{table}

Validation of a model is achieved when it is demonstrated that the model is free from requirements errors and reflects the stakeholders' wishes adequately. This can be done using several techniques, e.g., animation, review or walk-through. The most common way to validate a specification in the Event-B method is to animate the specification by invoking its operation semantics to inspect its behavior. We create behavioral scenarios and execute them. It is then examined whether the specification contains the desired functionality or not.

For model checking and animation of our specification, we have used the ProB tool \cite{leuschel03a} that supports automated consistency checking of Event-B machines via constraint solving techniques. Both model checking and animation using ProB worked very well. 

The ProB tool assisted us in finding potential invariant problems and their improvement by generating counterexamples whenever it discovered an invariant violation. It also helped us proving the deadlock freedom property of the model. ProB may also help in improving invariant expression by providing hints for strengthening invariants each time an invariant is modified or a new PO is generated by the Rodin platform. 

For animation, we created behavioral scenarios and executed them accordingly. We mainly demonstrated that the system is behaving as per expectations, no unintended path of executions is permissible and there is no violation of safety conditions by the specification. The resulting animation was easy to follow, especially for non-technical stakeholders.

As a matter of fact, we corrected more errors during specification modeling and reviewing than during discharging POs and animation.

\section{Lessons learned}
\label{sec:lessons}

During the model development exercise, we made certain experiences which are as follows:

\begin{itemize}

\item {\bf Formal models provide a consistent and complete repository of requirements}

The information presented in this paper as case study requirements does not possess a one-to-one mapping from the requirements document to the requirements specification. In fact, the data related to requirements was spread across several documents. For example, the alarm numbers were not explicitly stated in the requirements. Mining the relevant data from these documents is a time-consuming and tricky task. The broken or missing links may sometimes lead to incoherent information that may impact the correctness of the model. However, one of the advantages of the current modeling exercise is also to provide a repository of adequate and consistent requirements  that will positively impact the development of software.

\item {\bf Technical details impact the intelligibility of specifications}

The original purpose of formal specifications is to model and analyze requirements and design decisions in a way that leads to their systematic transformation into correct software. However, during the specification phase, sometimes we need to introduce additional constraints that are necessary to discharge POs but are not part of the original requirements document. Such technical elements impede the understandability of specifications for non-technical stakeholders. The practice of classification of axioms as described in Section \ref{subsec:strategy} not only increases the intelligibility of a specification but also helps distilling software requirements from technical constraints. The same procedure can be adopted for specifying machines. The guidelines proposed by Kossak et al. \cite{mashkoor14a} for writing understandable formal specifications by using proper naming conventions and structuring also help rendering specifications intelligible. 

\item {\bf There is no standard recipe for formal modeling}

Formal modeling is an overly complex engineering task that cannot be solved by applying some pre-cooked recipe \cite{su14a}. 
Different formal developments may require different modeling solutions. Traditionally, refinement-based development approaches follow a waterfall-like development structure where requirements are added to the model in a linear sequence. However, this practice does not suit well for large-scale model development \cite{mashkoor15b}. Therefore, diverging directions for model development are required. We tried multiple modeling approaches for specification development. They are as follows: 

\begin{itemize}

\item {\bf Linear sequence development} 

We first tried the traditional linear sequence approach. We started with one initial requirement and then continued enriching the model by introducing one requirement per refinement level. As expected, the model started becoming complex, proofs started becoming complicated and time consuming, and creating and running animation scenarios started becoming tedious with every refinement step. After six refinement steps, we reached a point where we decided that the model needs to be split. This is due to complex interactions involved between every newly introduced invariant and already existing events. 

\item {\bf Decomposable model development} 

The good thing about Event-B is that it provides means to decompose a large model either based on shared variables (A-Style) \cite{abrial07a} or shared events (B-Style) \cite{butler09a}. The former approach decomposes a model in such a way that sub-models can contain shared variables.  However, the shared variables cannot be refined. For instance, a model $M = \{v1, v2, v3\}$ can be split into $M1 = \{v1, v2\}$ and $ M2 = \{v2, v3\}$ where $v2$ is shared. The latter approach decomposes the model in such a way that sub-models have only distinct variables. For instance, a model $M = \{v1, v2, v3\}$ can be split into $M1 = \{v1\}$ and $ M2 = \{v2, v3\}$ where no variable is shared.

When we tried to decompose the dialysis machine model, we found out that the variable {\tt alarm} that is integral to the whole model cannot be present and further refined in every sub-model at the same time. So neither A-Style decomposition (although the variable {\tt alarm} would be present in every sub-model but could not be refined) nor B-Style decomposition (the variable {\tt alarm} would become part of only one sub-model) work in our case.

\item {\bf Atomic structure development}

The next solution that we tried was the atomic modeling of components; we modeled each component independently of the other components. This approach works well so far. The modeling is easy, proofs are straight forward and animation is quick. This is also the formalization that has been presented in this paper as the case study. The problem, however, with this approach is that currently we do not know how the components will interact with each other when they are plugged together. This requires further investigation.  
\end{itemize}
\end{itemize}

\section{Related work}
\label{sec:related}

Like other safety-critical systems, medical devices also benefit from formal methods. The use of formal methods has been in place for the development of various health-care products for a long time. Several decades ago, Hewlett Packard used HP-SL \cite{bear91a} to enhance the quality of a range of their cardiac care products \cite{bowen93a}. The University of Washington used the Z method \cite{spivey88a} for the development of a computer control system of a cyclotron and treatment facility that provides particle beams for cancer treatment \cite{jacky90a}.

In recent years, the use of formal methods is escalating for the development of software-intensive medical systems. For example, Osaiweran et al. \cite{osaiweran13a} use the formal Analytical Software Design (ASD) \cite{broadfoot05a} approach for developing the power control service of an interventional X-ray system. Jiant et al. \cite{jiang10a} present a methodology based on timed automata to extract timing properties of a heart that can be used for the verification and validation of implantable cardiac devices. Tuan et al. \cite{tuan10a} provide a solution for the pacemaker challenge using the model checker PAT (Process Analysis Toolkit) \cite{sun08a}. M\'{e}ry et al. \cite{mery13a} and Macedo et al. \cite{macedo08a} present a model of pacemakers in Event-B and VDM \cite{jones90a} respectively. 

One of the medical devices relatively close to hemodialysis  machines is an infusion pump. It is primarily responsible for delivering fluids, such as nutrients and medications, into a patient's body in controlled amounts. Arney et al. \cite{arney07a} present a reference model of PCA (Patient Control Analgesia) infusion pumps and test the model for structural and safety properties, Jos\'{e} et al. \cite{campos11a} present a formal model in MAL (Modal Action Logic) \cite{campos08a} that helps compare different infusion devices and their provided functionalities, and Bowen et al. \cite{bowen13a} use the ProZ model checker \cite{plagge07a} to test various safety properties of infusion pumps. 

The formal basis for medical software components development we have used in this paper is shared with aforementioned works. However, apart from the work of M\'{e}ry et al. \cite{mery13a}, the verification and validation activities are better integrated into our proposed development process as compared to others. We cover a multitude of model analysis activities, e.g., model checking, model review, and animation, that give us a grasp on the notion of correctness far better than approaches which are comprised of only a subset of analysis techniques we have employed. The work of M\'{e}ry et al. \cite{mery13a}, though based on Event-B, is still different than our work because they use the refinement chart approach for model development. It is a graphical modeling technique that provides a view of different subsystems offering assistance in their later integration into a single system. In contrast, our work is based on (different) conventional modeling strategies, i.e., linear sequential, decomposable and atomicity, whose semantics are well-defined and whose efficacy has been proven by several industrial success stories such as \cite{behm99a}, \cite{badeau05a}, \cite{iliasov13a}. The system under development is another difference; they work on pacemaker systems and our work is related to hemodialysis machines.

According to the best of our knowledge, this is the first instance of application of formal methods for the modeling and analysis of AMDs responsible for renal replacement therapy such as dialysis machines. We believe that our specification can act as a reference model that will inspire and facilitate manufacturers of such systems to adopt the formal paradigm for the safe and trustworthy development of variants of this domain.    

\section{Conclusion and future work}
\label{sec:conclusion}
This paper addresses the formal development of safety-critical software components embedded in AMDs. Ethics, as well as the necessity to comply with standards and regulations, make it imperative to follow an approach that helps in analyzing, specifying, implementing and testing such devices. 

In this paper, we also report on a case study of model-driven development of a hemodialysis machine, an instance of AMDs. Our conclusion regarding the two research questions defined in Section \ref{sec:design} is as follows:

\begin{enumerate}
\item Is the formal model-based approach sufficient for modeling all elements of a complex AMD? \\
The answer is Yes! Our employed approach successfully enabled us to specify and analyze various critical components of hemodialysis machines at different abstraction levels. The formal Event-B method supported by the Eclipse-based open source Rodin tool has also lent itself to the development of such systems. We have found Event-B an adequate method for the modeling and analysis of critical medical devices. Its refinement principles, and verification and validation mechanisms, provide all the elements that are necessary for the safe development of AMDs. 

\item What are advantages and challenges associated with this approach?\\
The apparent advantage of this approach is that we were able to ensure that errors and omissions in requirements are detected and corrected close to the point of their introduction. We were also successful in integrating non-technical stakeholders in the earlier phases of development cycle by showing them model animations and recording their feedback. The resulting formal model also provided a repository of adequate and consistent requirements that positively impacted the development of software. Last but not least, we found the Event-B notation relatively easy to learn and use. 

However, we also faced several challenges. For example, sophisticated tools and elaborated guidelines for managing the complexity of growing models by decomposition are missing. There is no implicit notion of time in Event-B, that is necessary for an elegant expression of timing properties which play a very important role in medical devices. Currently, we resort to ProB for proving temporal properties of the system. In our opinion, a standard and more natural way is required to specify and prove that temporal properties of the system are preserved by Event-B refinements. Finally, a tool that is able to generate {\it ready-to-deploy} machine code from formal models is also missing. Currently available Event-B supported tools for this purpose are clearly insufficient to produce code that can be deployed on hemodialysis machines without any further human intervention.   
\end{enumerate}

We are encouraged to proceed with the further development of components of hemodialysis machines. In future, we also plan to research on the transformation of requirements models into {\it ready-to-deploy} code artifacts.

\bibliographystyle{splncs}

\end{document}